# Generalized Lorentz law and the force of radiation on magnetic dielectrics

Masud Mansuripur

College of Optical Sciences, The University of Arizona, Tucson, Arizona 85721

**Abstract.** The macroscopic equations of Maxwell combined with a generalized form of the Lorentz law are a complete and consistent set; not only are these five equations fully compatible with the special theory of relativity, they also conform with the conservation laws of energy, momentum, and angular momentum. The linear momentum density associated with the electromagnetic field is $\boldsymbol{p}_{EM}(\boldsymbol{r},t) = \boldsymbol{E}(\boldsymbol{r},t) \times \boldsymbol{H}(\boldsymbol{r},t)/c^2$, whether the field is in vacuum or in a ponderable medium. [Homogeneous, linear, isotropic media are typically specified by their electric and magnetic permeabilities $\varepsilon_o \varepsilon(\omega)$ and $\mu_o \mu(\omega)$.] The electromagnetic momentum residing in a ponderable medium is often referred to as Abraham momentum. When an electromagnetic wave enters a medium, say, from the free space, it brings in Abraham momentum at a rate determined by the density distribution $\boldsymbol{p}_{EM}(\boldsymbol{r},t)$, which spreads within the medium with the light's group velocity. The balance of the incident, reflected, and transmitted (electromagnetic) momenta is subsequently transferred to the medium as mechanical force in accordance with Newton's second law. The mechanical force of the radiation field on the medium may also be calculated by a straightforward application of the generalized form of the Lorentz law. The fact that these two methods of force calculation yield identical results is the basis of our claim that the equations of electrodynamics (Maxwell + Lorentz) comply with the momentum conservation law. When applying the Lorentz law, one must take care to properly account for the effects of material dispersion and absorption, discontinuities at material boundaries, and finite beam dimensions. This paper demonstrates some of the issues involved in such calculations of the electromagnetic force in magnetic dielectric media.

**1. Introduction.** Maxwell's macroscopic equations in conjunction with a generalized form of the Lorentz law are consistent with the laws of conservation of energy, momentum, and angular momentum. In recent publications we have demonstrated this consistency by showing that, when a beam of light enters a magnetic dielectric, a fraction of the incident linear (or angular) momentum pours into the medium at a rate determined by the Abraham momentum density, $\boldsymbol{E} \times \boldsymbol{H}/c^2$, as well as by the group velocity $V_g$ of the electromagnetic field. The balance of the incident, reflected, and transmitted linear (angular) momenta is subsequently transferred to the medium as force (torque), usually at the leading edge of the beam, which propagates through the medium with velocity $V_g$. When expressing force, torque, and momentum densities, our analysis generally equates electromagnetic momentum with Abraham momentum [1], and distinguishes the phase refractive index $n_p$ from the group refractive index $n_g$.

Standard textbooks on electromagnetism tend to treat the macroscopic Maxwell's equations as somehow inferior to their microscopic counterparts [2,3]. This is due to the fact that, for real materials, polarization and magnetization densities **P** and **M** are defined as averages over small volumes that must nevertheless contain a large number of atomic dipoles. Consequently, the macroscopic **E**, **D**, **H** and **B** fields are regarded as spatial averages of the "actual" fields; without averaging, these fields would be wildly fluctuating on the scale of atomic dimensions. (The actual fields, of course, are assumed to be well-defined at all points in space and time.) There is also a tendency to elevate the **E** and **B** fields to the status of "fundamental," while treating **D** and **H** as secondary in importance. This is an unfortunate state of affairs, considering that the macroscopic equations of Maxwell are a complete and self-consistent set, provided that the fields are treated as precisely-defined mathematical entities, i.e., without attempting to associate **P** and **M** with the properties of real materials. Stated differently, if material media consisted of dense collections of point dipoles, then any volume of the material, no matter how small, would contain an infinite number of such dipoles, eliminating thereby the need for the introduction of macroscopic averages into Maxwell's equations. Also, since in their simplest form, the macroscopic equations contain all four of the **E**, **D**, **H**, **B** fields, one should perhaps resist the temptation to designate some of these as more fundamental than others. Tellegen [4] regards these four fields as equally important, a point of view with which we tend to agree.

Constitutive relations equate **P** with $\boldsymbol{D} - \varepsilon_o \boldsymbol{E}$ and **M** with $\boldsymbol{B} - \mu_o \boldsymbol{H}$, thus allowing **P** and **M** to be designated as secondary fields. Electric and magnetic energy densities may now be written as $\boldsymbol{E} \cdot \boldsymbol{D}$ and $\boldsymbol{H} \cdot \boldsymbol{B}$, respectively, and the Poynting vector can be expressed as $\boldsymbol{S} = \boldsymbol{E} \times \boldsymbol{H}$, without the need to explain away the appearance of the "derived" fields **D** and **H** in the expression of a most fundamental physical entity. (Note that, in deriving Poynting's theorem, the assumed rate of change of energy density is $\partial \mathcal{E}/\partial t = \boldsymbol{E} \cdot \boldsymbol{J}_{\text{free}} + \boldsymbol{E} \cdot \partial \boldsymbol{D}/\partial t + \boldsymbol{H} \cdot \partial \boldsymbol{B}/\partial t$. This, in fact, is the only postulate of the classical theory concerning electromagnetic energy.)

The fifth fundamental equation of the classical theory, the Lorentz law of force $\boldsymbol{F} = q(\boldsymbol{E} + \boldsymbol{V} \times \boldsymbol{B})$, expresses the force experienced by a particle of charge $q$ moving with velocity $\boldsymbol{V}$ through the electromagnetic field [2,3]. It is fairly straightforward to derive from this law the force and torque exerted on an electric dipole $\boldsymbol{p}$ (or the corresponding densities exerted on the polarization $\boldsymbol{P}$). However, the Lorentz law is silent on the question of force/torque experienced by a magnetic dipole $\boldsymbol{m}$ in the presence of an electromagnetic field. Traditionally, magnetic dipoles have been treated as Amperian current loops, and the force/torque exerted upon them have been derived from the standard Lorentz law by considering the loop's current as arising from circulating electric charges. The problem with this approach is that, when examining the propagation of electromagnetic waves through magnetic media, one finds that linear and angular momenta are *not* conserved. Shockley [5] has famously called attention to the problem of "hidden" momentum within magnetic materials. Fortunately, it is possible to extend the Lorentz law to include the electromagnetic forces on both electric and magnetic dipoles in a way that is consistent with the conservation of energy, momentum, and angular momentum. This extension of the Lorentz law has been attempted a few times during the past forty years, each time from a different perspective, but always resulting in essentially the same generalized form of the force law [5-11]. It is now possible to claim that we finally possess a generalized Lorentz law which, in conjunction with Maxwell's macroscopic equations, is fully consistent with the conservation laws as well as with the principles of special relativity.

The goal of the present paper is to demonstrate the consistency of the generalized Lorentz law with the law of momentum conservation in two special cases. We confine our attention to homogeneous, linear, isotropic media specified by their relative permittivity $\varepsilon(\omega)$ and permeability $\mu(\omega)$. The first example, pertaining to a thin, parallel-plate slab discussed in Section 3, is applicable to transparent media, whose $(\varepsilon, \mu)$ are real-valued, as well as to absorbing media, where at least one of these parameters is complex. In the second example, discussed in Section 4, we investigate a case involving total internal reflection within a transparent prism made up of a magnetic dielectric material.

**2. Force and torque of the electromagnetic field on electric and magnetic dipoles.** In a recent publication [11] we derived the following generalized expressions for the Lorentz force and torque densities in a homogeneous, linear, isotropic medium specified by its $\mu$ and $\varepsilon$ parameters:

$$\boldsymbol{F}_1(\boldsymbol{r},t) = (\boldsymbol{P} \cdot \nabla)\boldsymbol{E} + (\boldsymbol{M} \cdot \nabla)\boldsymbol{H} + (\partial \boldsymbol{P}/\partial t) \times \mu_o \boldsymbol{H} - (\partial \boldsymbol{M}/\partial t) \times \varepsilon_o \boldsymbol{E}, \tag{1a}$$

$$\boldsymbol{T}_1(\boldsymbol{r},t) = \boldsymbol{r} \times \boldsymbol{F}_1(\boldsymbol{r},t) + \boldsymbol{P}(\boldsymbol{r},t) \times \boldsymbol{E}(\boldsymbol{r},t) + \boldsymbol{M}(\boldsymbol{r},t) \times \boldsymbol{H}(\boldsymbol{r},t). \tag{1b}$$

Similar expressions have been derived by others (see, for example, Hansen and Yaghjian [9]). Our focus, however, has been the generalization of the Lorentz law in a way that is consistent with Maxwell's equations, with the principles of special relativity, and with the laws of conservation of energy and momentum. In conjunction with Eqs. (1), Maxwell's equations in the MKSA system of units are:

$$\nabla \cdot \boldsymbol{D} = \rho_{\text{free}}; \qquad \nabla \times \boldsymbol{H} = \boldsymbol{J}_{\text{free}} + \partial \boldsymbol{D}/\partial t; \qquad \nabla \times \boldsymbol{E} = -\partial \boldsymbol{B}/\partial t; \qquad \nabla \cdot \boldsymbol{B} = 0. \tag{2}$$

In these equations, the electric displacement $\boldsymbol{D}$ and the magnetic induction $\boldsymbol{B}$ are related to the polarization density $\boldsymbol{P}$ and the magnetization density $\boldsymbol{M}$ via the constitutive relations:

$$\boldsymbol{D} = \varepsilon_o \boldsymbol{E} + \boldsymbol{P} = \varepsilon_o(1 + \chi_e)\boldsymbol{E} = \varepsilon_o \varepsilon \boldsymbol{E}; \qquad \boldsymbol{B} = \mu_o \boldsymbol{H} + \boldsymbol{M} = \mu_o(1 + \chi_m)\boldsymbol{H} = \mu_o \mu \boldsymbol{H}. \tag{3}$$

In what follows, the medium will be assumed to have neither free charges nor free currents (i.e., $\rho_{\text{free}} = 0$, $\boldsymbol{J}_{\text{free}} = 0$). Our linear isotropic media will be assumed to be fully specified by their permittivity $\varepsilon = \varepsilon' + i\varepsilon''$ and permeability $\mu = \mu' + i\mu''$. Any loss of energy in such media will be associated with $\varepsilon''$ and $\mu''$, which, by convention, are $\geq 0$. The real parts of $\varepsilon$ and $\mu$, however, may be either positive or negative. In particular, $\varepsilon' < 0$ and $\mu' < 0$ in negative-index media.

In [11] and elsewhere [12-20], we have considered an alternative formulation of the generalized Lorentz law, where bound electric and magnetic charge densities $\rho_e = -\nabla \cdot \boldsymbol{P}$ and $\rho_m = -\nabla \cdot \boldsymbol{M}$ directly experience the force of the $\boldsymbol{E}$ and $\boldsymbol{H}$ fields. The alternative formulas are

$$\boldsymbol{F}_2(\boldsymbol{r},t) = -(\nabla \cdot \boldsymbol{P})\boldsymbol{E} - (\nabla \cdot \boldsymbol{M})\boldsymbol{H} + (\partial \boldsymbol{P}/\partial t) \times \mu_o \boldsymbol{H} - (\partial \boldsymbol{M}/\partial t) \times \varepsilon_o \boldsymbol{E}, \tag{4a}$$

$$\boldsymbol{T}_2(\boldsymbol{r},t) = \boldsymbol{r} \times \boldsymbol{F}_2(\boldsymbol{r},t). \tag{4b}$$

As far as the total force (or torque) exerted on a given volume of material is concerned, Eqs. (1) and (4) can be shown to yield identical results provided that the boundaries are properly treated [21, 22]. The force (or torque) distribution throughout a given volume, of course, will depend on which formulation is used, but when integrated over the volume of interest, the two distributions always yield identical values for the total force (or torque). The proof of equivalence of total force (and total torque) for the two formulations was originally given by Barnett and Loudon in [21,22]. Subsequently, we extended their proof to cover the case of objects immersed in a liquid [23,24]. In our proof, we stated that $\boldsymbol{P}\times\boldsymbol{E}$ (and, by analogy, $\boldsymbol{M}\times\boldsymbol{H}$) will be zero in isotropic media and, therefore, the additional terms in Eq. (1b) need not be considered. This statement, while valid in some cases, is generally incorrect. In other words, for the total torques in the two formulations to be identical, Eq. (1b) must retain the $\boldsymbol{P}\times\boldsymbol{E}$ and $\boldsymbol{M}\times\boldsymbol{H}$ contributions.

Using simple examples that are amenable to exact analysis, we have shown in previous publications [11-20] that Eqs. (1-4) lead to a precise balance of linear and angular momenta when all relevant forces, especially those at the boundaries, are properly taken into account. The concern of the present paper is to provide further evidence in support of the generalized Lorentz law by extending our calculations to more complex situations.

**3. Radiation pressure on a dipole sheet**. Figure 1 shows a thin slab of a material specified by its $(\varepsilon,\mu)$ parameters. A linearly polarized plane-wave is normally incident on this slab from the left-hand side. The reflected and transmitted amplitudes are denoted by $\rho$ and $\tau$, respectively. Here we are interested in the behavior of a thin slab, i.e., when $d\to 0$. We denote by $A$ and $B$ the amplitudes of the two plane-waves that propagate to the right and left inside the slab. Matching the boundary conditions yields the following expressions for the reflection and transmission coefficients, the fields $E_x(z)$ and $H_y(z)$ inside the slab, and the field amplitudes $E_c$ and $H_c$ at the center of the slab (where $z=\tfrac{1}{2}d$):

$$\rho = \frac{\tfrac{1}{2}(\mu-\varepsilon)\sin(k_o\sqrt{\mu\varepsilon}\,d)}{\tfrac{1}{2}(\mu+\varepsilon)\sin(k_o\sqrt{\mu\varepsilon}\,d)+i\sqrt{\mu\varepsilon}\cos(k_o\sqrt{\mu\varepsilon}\,d)} \approx -\tfrac{1}{2}i(\mu-\varepsilon)k_o d + \tfrac{1}{4}(\mu^2-\varepsilon^2)k_o^2 d^2, \tag{5a}$$

$$\tau\exp(ik_o d) = \frac{i\sqrt{\mu\varepsilon}}{\tfrac{1}{2}(\mu+\varepsilon)\sin(k_o\sqrt{\mu\varepsilon}\,d)+i\sqrt{\mu\varepsilon}\cos(k_o\sqrt{\mu\varepsilon}\,d)} \approx 1 + \tfrac{1}{2}i(\mu+\varepsilon)k_o d - \tfrac{1}{4}(\mu^2+\varepsilon^2)k_o^2 d^2, \tag{5b}$$

$$E_x(z)=E_o\frac{\mu\sin[k_o\sqrt{\mu\varepsilon}(d-z)]+i\sqrt{\mu\varepsilon}\cos[k_o\sqrt{\mu\varepsilon}(d-z)]}{\tfrac{1}{2}(\mu+\varepsilon)\sin(k_o\sqrt{\mu\varepsilon}\,d)+i\sqrt{\mu\varepsilon}\cos(k_o\sqrt{\mu\varepsilon}\,d)}; \qquad 0\le z\le d, \tag{5c}$$

$$H_y(z)=Z_o^{-1}E_o\frac{\varepsilon\sin[k_o\sqrt{\mu\varepsilon}(d-z)]+i\sqrt{\mu\varepsilon}\cos[k_o\sqrt{\mu\varepsilon}(d-z)]}{\tfrac{1}{2}(\mu+\varepsilon)\sin(k_o\sqrt{\mu\varepsilon}\,d)+i\sqrt{\mu\varepsilon}\cos(k_o\sqrt{\mu\varepsilon}\,d)}; \qquad 0\le z\le d, \tag{5d}$$

$$E_c = E_x(z=\tfrac{1}{2}d) \approx [1 + \tfrac{1}{2}i\varepsilon k_o d + \tfrac{1}{8}(\mu\varepsilon - 2\varepsilon^2)k_o^2 d^2]E_o, \tag{5e}$$

$$H_c = H_y(z=\tfrac{1}{2}d) \approx [1 + \tfrac{1}{2}i\mu k_o d + \tfrac{1}{8}(\mu\varepsilon - 2\mu^2)k_o^2 d^2]Z_o^{-1}E_o. \tag{5f}$$

In the above equations $k_o=2\pi/\lambda_o=\omega/c$ is the wave-number, $c=1/\sqrt{\mu_o\varepsilon_o}$ is the speed of light in vacuum, and $Z_o=\sqrt{\mu_o/\varepsilon_o}\approx 377\Omega$ is the impedance of the free space. It is now easy to determine the time-averaged force $\langle F_z\rangle$ on the slab using either the conservation of momentum in the free space, namely, $\langle F_z\rangle = \tfrac{1}{2}\varepsilon_o E_o^2(1+|\rho|^2-|\tau^2|)$, or by direct computation from Eq. (1a) using the mid-point fields $E_c$ and $H_c$. Either way we find:

$$\langle F_z\rangle \approx \tfrac{1}{2}\varepsilon_o E_o^2[(\mu''+\varepsilon'')k_o d - \tfrac{1}{2}(\mu''+\varepsilon'')^2 k_o^2 d^2 + \tfrac{1}{2}(\mu'-\varepsilon')^2 k_o^2 d^2]. \tag{6}$$

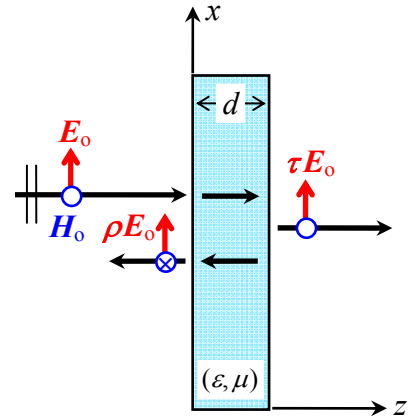

**Figure 1**. A slab of thickness $d$ and material parameters $(\varepsilon,\mu)$ is illuminated at normal incidence with a linearly polarized plane-wave. The amplitude reflection and transmission coefficients are denoted by $\rho$ and $\tau$, respectively. In the limit of small $d$, the slab resembles a single sheet of electric and magnetic dipoles excited by the incident wave. The excited dipoles radiate in both directions ($\pm z$), with the wave that propagates to the left accounting for the reflected light, and the wave that propagates to the right interfering with the (un-attenuated) incident beam to produce the transmitted light.

Note that the part of $\langle F_z \rangle$ that is due to absorption in the material (i.e., the first two terms of Eq. (6), which contain $\mu''$ and $\varepsilon''$) is proportional to the slab thickness $d$ in the limit of a thin slab. However, in the absence of absorption, the first two terms of Eq. (6) drop out, leaving only the third term, which is proportional to $d^2$.

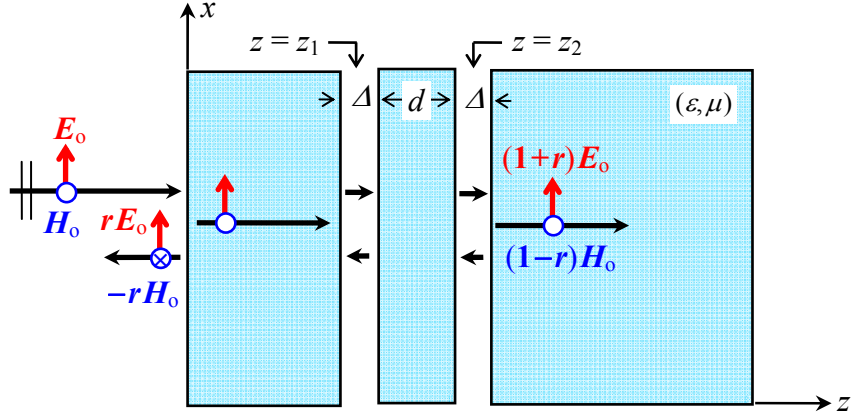

**Figure 2**. A transparent, semi-infinite slab specified by its material parameters $(\varepsilon, \mu)$ is illuminated at normal incidence with a linearly polarized plane-wave. The Fresnel reflection coefficient at the front facet is denoted by $r$. Any internal slab of finite thickness $d$, although an integral part of the medium, may be imagined to be surrounded on both sides by narrow gaps of width $\Delta \ll \lambda_o$. The symmetry of the incident, reflected, and transmitted momenta on the two sides of the embedded slab, irrespective of its thickness $d$, ensures that the net force of radiation on the embedded slab is zero.

Shown in Fig. 2 is a transparent, semi-infinite slab of material parameters $(\varepsilon, \mu)$, illuminated at normal incidence with a linearly polarized plane-wave. The amplitude reflection coefficient at the front facet is denoted by $r$. Inside the slab the field amplitudes are

$$E_x(z, t) = (1+r)E_o \exp(\mathrm{i}k_o\sqrt{\mu\varepsilon}\,z - \mathrm{i}\omega t), \tag{7a}$$

$$H_y(z, t) = (1-r)H_o \exp(\mathrm{i}k_o\sqrt{\mu\varepsilon}\,z - \mathrm{i}\omega t). \tag{7b}$$

Any internal segment of finite thickness $d$, although an integral part of the medium, can be imagined to be surrounded by narrow gaps of width $\Delta \ll \lambda_o$ on either side. Within the gap centered at $z = z_1$, there exist counter-propagating plane-waves with $E$-field amplitudes $E_o\exp(\mathrm{i}k_o\sqrt{\mu\varepsilon}\,z_1)$ and $rE_o\exp(\mathrm{i}k_o\sqrt{\mu\varepsilon}\,z_1)$. Similarly, the $E$-field amplitudes of the two plane-waves residing within the gap at $z = z_2$ are $E_o\exp(\mathrm{i}k_o\sqrt{\mu\varepsilon}\,z_2)$ and $rE_o\exp(\mathrm{i}k_o\sqrt{\mu\varepsilon}\,z_2)$. Since the two beams in each gap are counter-propagating, their $E$-field amplitudes are added together, while their $H$-field amplitudes are subtracted from each other to yield the total gap field; this yields the same $E$- and $H$-fields that are known to reside within the semi-infinite slab at each point along the $z$-axis. [One can readily show, with the aid of Eqs. (5a, 5b), that the gap fields are consistent with the reflection and transmission coefficients $\rho$ and $\tau$ of the slab of thickness $d$ depicted in Fig. 1.] The symmetry of the incident, reflected, and transmitted momenta on the two sides of the embedded slab reveals that such (embedded) sheets of transparent material should experience no radiation forces, irrespective of their thickness $d$. In the limit when $d \to 0$, the phase difference between $E_c$ and $H_c$ at the center of the dipole sheet vanishes. Hence, due to incidence from both sides of the sheet, the net radiation force on embedded dipoles will be zero. This argument applies to surface dipoles as well as to any other layer of dipoles embedded within the semi-infinite slab.

**4. Total internal reflection within a transparent magnetic dielectric prism**. Figure 3 shows a monochromatic, linearly-polarized, finite-diameter beam incident at an oblique angle $\theta$ at the base of a transparent, homogeneous, linear, isotropic prism. In this two-dimensional problem the beam's width is finite only in the $xz$-plane, its profile being uniform along the $y$-axis from $-\infty$ to $\infty$. Inside the prism, whose material parameters are $(\varepsilon, \mu)$, the electric and magnetic fields may be expressed as the following Fourier integrals:

$$\boldsymbol{E}(\boldsymbol{r}, t) = \tfrac{1}{2}\left\{\int \boldsymbol{\mathcal{E}}(k_x)\exp[\mathrm{i}(k_x x + k_z z - \omega t)]\,\mathrm{d}k_x + \int \boldsymbol{\mathcal{E}}^*(k_x)\exp[-\mathrm{i}(k_x x + k_z z - \omega t)]\,\mathrm{d}k_x\right\}, \tag{8a}$$

$$\boldsymbol{H}(\boldsymbol{r}, t) = \tfrac{1}{2}\left\{\int \boldsymbol{\mathcal{H}}(k_x)\exp[\mathrm{i}(k_x x + k_z z - \omega t)]\,\mathrm{d}k_x + \int \boldsymbol{\mathcal{H}}^*(k_x)\exp[-\mathrm{i}(k_x x + k_z z - \omega t)]\,\mathrm{d}k_x\right\}. \tag{8b}$$

Here $\omega \geq 0$ is the fixed oscillation frequency, the range of integration over $k_x$ is a narrow band in the vicinity of $k_{xo} = (\omega/c)\sqrt{\mu\varepsilon}\sin\theta$, and $k_z = (\omega/c)\sqrt{\mu\varepsilon - (ck_x/\omega)^2}$. The magnetic field components of the plane-waves constituting the incident beam are given by Maxwell's third equation, $\nabla \times \boldsymbol{E} = -\partial \boldsymbol{B}/\partial t$, as follows:

$$k_y\mathcal{E}_z - k_z\mathcal{E}_y = \omega\mu_o\mu\mathcal{H}_x, \quad k_z\mathcal{E}_x - k_x\mathcal{E}_z = \omega\mu_o\mu\mathcal{H}_y, \quad k_x\mathcal{E}_y - k_y\mathcal{E}_x = \omega\mu_o\mu\mathcal{H}_z. \tag{9}$$

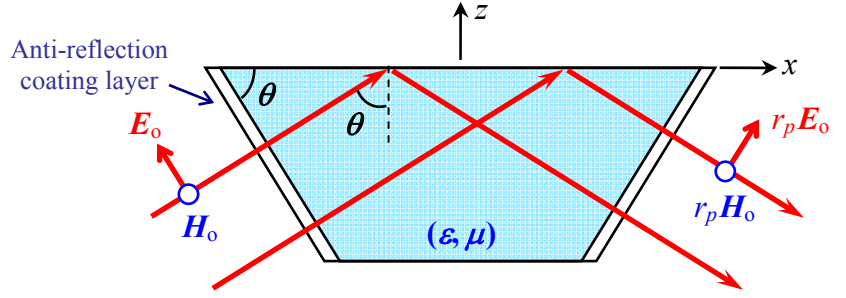

**Figure 3**. A finite-diameter beam of light enters a magnetic dielectric prism whose material parameters are $(\varepsilon,\mu)$. The beam enters at normal incidence to the entrance facet, is totally reflected at the upper surface, then exits at normal incidence to the exit facet. The entrance and exit facets are anti-reflection coated, so that the beam in its entirety enters and exits the prism. The beam is p-polarized (i.e., TM), with its field components being $(E_x, E_z, H_y)$.

For a TM-polarized wave the force density in accordance with Eq. (1a) is given by

$$\boldsymbol{F}(\boldsymbol{r},t) = P_x(\partial \boldsymbol{E}/\partial x) + P_z(\partial \boldsymbol{E}/\partial z) + M_y(\partial \boldsymbol{H}/\partial y) + (\partial \boldsymbol{P}/\partial t) \times \mu_0 \boldsymbol{H} - (\partial \boldsymbol{M}/\partial t) \times \varepsilon_0 \boldsymbol{E}. \tag{10}$$

Substituting from Eqs. (8) into Eq. (10), multiplying through, then time-averaging to eliminate terms with $\pm 2\omega$ in their exponents, yields

$$\langle \boldsymbol{F}(\boldsymbol{r},t)\rangle = \tfrac{1}{4}\mathrm{i}\,\varepsilon_0(\varepsilon-1)\Big\{\iint [k_x \mathcal{E}(k_x)\mathcal{E}_x^*(k_x') - k_x' \mathcal{E}_x(k_x)\mathcal{E}^*(k_x')] \exp[\mathrm{i}(k_x-k_x')x]\exp[\mathrm{i}(k_z-k_z')z]\,\mathrm{d}k_x \mathrm{d}k_x'$$

$$+ \iint [k_z \mathcal{E}(k_x)\mathcal{E}_z^*(k_x') - k_z' \mathcal{E}(k_x)\mathcal{E}_z^*(k_x')]\exp[\mathrm{i}(k_x-k_x')x]\exp[\mathrm{i}(k_z-k_z')z]\,\mathrm{d}k_x \mathrm{d}k_x'\Big\}$$

$$- \tfrac{1}{4}\mathrm{i}\,\mu_0 \varepsilon_0(\varepsilon-1)\omega \iint [\mathcal{E}(k_x)\times \mathcal{H}^*(k_x') - \mathcal{E}^*(k_x')\times \mathcal{H}(k_x)]\exp[\mathrm{i}(k_x-k_x')x]\exp[\mathrm{i}(k_z-k_z')z]\,\mathrm{d}k_x \mathrm{d}k_x'$$

$$+ \tfrac{1}{4}\mathrm{i}\,\mu_0 \varepsilon_0(\mu-1)\omega \iint [\mathcal{H}(k_x)\times \mathcal{E}^*(k_x') - \mathcal{H}^*(k_x')\times \mathcal{E}(k_x)]\exp[\mathrm{i}(k_x-k_x')x]\exp[\mathrm{i}(k_z-k_z')z]\,\mathrm{d}k_x \mathrm{d}k_x'. \tag{11}$$

Combining similar terms, then replacing the $H$-field in terms of the $E$-field components as given by Eq. (9), we obtain

$$\langle \boldsymbol{F}\rangle = \tfrac{1}{4}\mathrm{i}\,\varepsilon_0(\varepsilon-1)\iint \Big\{[k_x\mathcal{E}_x(k_x)\mathcal{E}_x^*(k_x') - k_x'\mathcal{E}_x(k_x)\mathcal{E}_x^*(k_x') + k_z\mathcal{E}_x(k_x)\mathcal{E}_z^*(k_x') - k_z'\mathcal{E}_z(k_x)\mathcal{E}_x^*(k_x')]\hat{\boldsymbol{x}}$$

$$+ [k_x\mathcal{E}_z(k_x)\mathcal{E}_x^*(k_x') - k_x'\mathcal{E}_x(k_x)\mathcal{E}_z^*(k_x') + k_z\mathcal{E}_z(k_x)\mathcal{E}_z^*(k_x') - k_z'\mathcal{E}_z(k_x)\mathcal{E}_z^*(k_x')]\hat{\boldsymbol{z}}\Big\}$$

$$\times \exp[\mathrm{i}(k_x-k_x')x]\exp[\mathrm{i}(k_z-k_z')z]\,\mathrm{d}k_x \mathrm{d}k_x'$$

$$+ \tfrac{1}{4}\mathrm{i}\,\varepsilon_0[(\mu-\varepsilon)/\mu]\iint \Big\{[k_x'\mathcal{E}_z(k_x)\mathcal{E}_z^*(k_x') - k_x\mathcal{E}_z(k_x)\mathcal{E}_z^*(k_x') + k_z\mathcal{E}_x(k_x)\mathcal{E}_x^*(k_x') - k_z'\mathcal{E}_x(k_x)\mathcal{E}_x^*(k_x')]\hat{\boldsymbol{x}}$$

$$+ [k_z'\mathcal{E}_x(k_x)\mathcal{E}_x^*(k_x') - k_z\mathcal{E}_x(k_x)\mathcal{E}_x^*(k_x') - k_x'\mathcal{E}_x(k_x)\mathcal{E}_z^*(k_x') + k_x\mathcal{E}_z(k_x)\mathcal{E}_x^*(k_x')]\hat{\boldsymbol{z}}\Big\}$$

$$\times \exp[\mathrm{i}(k_x-k_x')x]\exp[\mathrm{i}(k_z-k_z')z]\,\mathrm{d}k_x \mathrm{d}k_x'. \tag{12}$$

Next, using Maxwell's first equation, $k_x\mathcal{E}_x + k_z\mathcal{E}_z = 0$, we further simplify the above force density expression and find

$$\langle \boldsymbol{F}\rangle = \tfrac{1}{4}\mathrm{i}\,\varepsilon_0 \iint \Big\{\big\{(\varepsilon-1)(k_x-k_x')\mathcal{E}_x(k_x)\mathcal{E}_x^*(k_x') + \{[(\varepsilon/\mu)-1](k_x-k_x') + [(\varepsilon/\mu)-\varepsilon][(k_z^2/k_x) - (k_z'^2/k_x')]\}\mathcal{E}_z(k_x)\mathcal{E}_z^*(k_x')\big\}\hat{\boldsymbol{x}}$$

$$+ \big\{[(\varepsilon/\mu)-1](k_z-k_z')\mathcal{E}_x(k_x)\mathcal{E}_x^*(k_x') + \{(\varepsilon-1)(k_z-k_z') - [(\varepsilon/\mu)-\varepsilon][(k_x'k_z/k_x) - (k_xk_z'/k_x')]\}\mathcal{E}_z(k_x)\mathcal{E}_z^*(k_x')\big\}\hat{\boldsymbol{z}}\Big\}$$

$$\times \exp[\mathrm{i}(k_x-k_x')x]\exp[\mathrm{i}(k_z-k_z')z]\,\mathrm{d}k_x \mathrm{d}k_x'. \tag{13}$$

To find the total force exerted by the incident beam inside the prism of Fig. 3, we integrate Eq. (13), first from $z(x) = -[h+(\tan\theta)x]$ to $0$, where $h$ is the height of the full prism at $x=0$, then over $x$ from $-\infty$ to $\infty$ to obtain:

$$\int_{-\infty}^{\infty}\int_{z(x)}^{0} \langle \boldsymbol{F}(x,z,t)\rangle\,\mathrm{d}z\,\mathrm{d}x$$

$$= \tfrac{1}{4}\varepsilon_0 \iint \Big\{\big\{(\varepsilon-1)(k_x-k_x')\mathcal{E}_x(k_x)\mathcal{E}_x^*(k_x') + \{[(\varepsilon/\mu)-1](k_x-k_x') + [(\varepsilon/\mu)-\varepsilon][(k_z^2/k_x) - (k_z'^2/k_x')]\}\mathcal{E}_z(k_x)\mathcal{E}_z^*(k_x')\big\}\hat{\boldsymbol{x}}$$

$$+ \big\{[(\varepsilon/\mu)-1](k_z-k_z')\mathcal{E}_x(k_x)\mathcal{E}_x^*(k_x') + \{(\varepsilon-1)(k_z-k_z') - [(\varepsilon/\mu)-\varepsilon][(k_x'k_z/k_x) - (k_xk_z'/k_x')]\}\mathcal{E}_z(k_x)\mathcal{E}_z^*(k_x')\big\}\hat{\boldsymbol{z}}\Big\}$$

$$\times (k_z-k_z')^{-1}\big\{\delta(k_x-k_x') - \exp[-\mathrm{i}(k_z-k_z')h]\,\delta[(k_x-k_x') - (k_z-k_z')\tan\theta]\big\}\,\mathrm{d}k_x \mathrm{d}k_x'. \tag{14}$$

The second $\delta$-function may be simplified by finding the derivative (with respect to $k_x$) of its argument in the vicinity of $k_x = k_x'$; this shows the second $\delta$-function to be equal to $[1+(k_x/k_z)\tan\theta]^{-1}\delta(k_x-k_x')$. The phase factor $\exp[-i(k_z-k_z')h]$ multiplying this $\delta$-function now disappears because the exponent vanishes at $k_x = k_x'$. This is a welcome development considering that, while the slope $\tan\theta$ of the prism's entrance facet plays an important role, the prism height $h$ has no relevance. The two $\delta$-functions now combine and are replaced with $\{\tan\theta/[\tan\theta+(k_z/k_x)]\}\delta(k_x-k_x')$. Integration over $k_x'$ means that the entire integrand must be evaluated at $k_x = k_x'$. To eliminate the singularity created by $(k_z-k_z')$ in the denominator, we expand the various coefficients around the point $k_x = k_x'$. This yields

$$k_z - k_z' \approx -(k_x/k_z)(k_x - k_x'); \quad (k_z^2/k_x) - (k_z'^2/k_x') \approx -[2+(k_z/k_x)^2](k_x - k_x'); \quad (k_x' k_z/k_x) - (k_x k_z'/k_x') \approx [1+2(k_z/k_x)^2](k_z - k_z'). \quad (15)$$

Substitution into Eq. (14) yields

$$\int_{-\infty}^{\infty}\int_{z(x)}^{0} \langle \boldsymbol{F}(x,z,t)\rangle\, \mathrm{d}z\, \mathrm{d}x$$

$$= \tfrac{1}{4}\varepsilon_0 \tan\theta \int [\tan\theta + (k_z/k_x)]^{-1} \left\{ (k_z/k_x)\{(1-\varepsilon)|\mathcal{E}_x(k_x)|^2 + \{1+(\varepsilon/\mu)-2\varepsilon+[(\varepsilon/\mu)-\varepsilon](k_z/k_x)^2]\}|\mathcal{E}_z(k_x)|^2\}\hat{\boldsymbol{x}} \right.$$

$$\left. + \{[(\varepsilon/\mu)-1]|\mathcal{E}_x(k_x)|^2 - \{1+(\varepsilon/\mu)-2\varepsilon+2[(\varepsilon/\mu)-\varepsilon](k_z/k_x)^2\}|\mathcal{E}_z(k_x)|^2\}\hat{\boldsymbol{z}} \right\} \mathrm{d}k_x. \quad (16)$$

In the limit of a wide beam having a narrow spectral distribution centered around $k_{xo} = (\omega/c)\sqrt{\mu\varepsilon}\sin\theta$, we treat $k_x/k_z \approx \tan\theta$ as a constant. Also, $|\mathcal{E}_x| \approx |\mathcal{E}|\cos\theta$ and $|\mathcal{E}_z| \approx |\mathcal{E}|\sin\theta$. The above integral thus simplifies as follows:

$$\int_{-\infty}^{\infty}\int_{z(x)}^{0} \langle \boldsymbol{F}(x,z,t)\rangle\, \mathrm{d}z\, \mathrm{d}x = \tfrac{1}{4}\varepsilon_0[(\varepsilon/\mu)-2\varepsilon+1]\left[\int_{-\infty}^{\infty}|\mathcal{E}(k_x)|^2 \mathrm{d}k_x\right]\sin\theta(\cos\theta\, \hat{\boldsymbol{x}} - \sin\theta\, \hat{\boldsymbol{z}}). \quad (17)$$

According to Parseval's theorem of Fourier analysis, the integral of $|\mathcal{E}|^2$ over $k_x$ is the same as that of $|\boldsymbol{E}|^2$ over the footprint of the beam along the $x$-axis. The factor $\sin\theta$ thus accounts for the excess length of the incident beam's right-hand sidewall. Therefore, what we have in Eq. (17) is the net radiation pressure at the side-walls of the incident beam of Fig. 3, in agreement with the corresponding formula derived in [11].

Next we evaluate the radiation pressure exerted on the prism by the reflected beam. Each plane-wave in the Fourier spectrum of this beam is multiplied by $r_p(k_x)$, the Fresnel reflection coefficient for a p-polarized plane-wave at oblique incidence at the prism's base. The derivation parallels that given above for the incident beam. In the end, since the square of the magnitudes of all field components are multiplied by $|r_p(k_x)|^2$, and since at total internal reflection this coefficient is unity, the force exerted by the reflected beam turns out to be the same as that given by Eq. (17) for the incident beam, except for a sign reversal of the $x$-component, which is expected from the symmetry of the problem.

There remains to evaluate the force due to interference between the incident and reflected beams in the region where the two beams overlap. The procedure is similar to that outlined in Eqs. (8-17) above, with the exception that $r_p(k_x)$ must now be used to define the reflected field amplitudes. The incident plus reflected $E$-field is thus written

$$\boldsymbol{E}(\boldsymbol{r},t) = \tfrac{1}{2}\left\{\int \mathcal{E}(k_x)\exp[\mathrm{i}(k_x x + k_z z - \omega t)]\mathrm{d}k_x + \int \mathcal{E}^*(k_x)\exp[-\mathrm{i}(k_x x + k_z z - \omega t)]\mathrm{d}k_x \right.$$

$$\left. + \int r_p(k_x)\mathcal{E}(k_x)\exp[\mathrm{i}(k_x x - k_z z - \omega t)]\mathrm{d}k_x + \int r_p^*(k_x)\mathcal{E}^*(k_x)\exp[-\mathrm{i}(k_x x - k_z z - \omega t)]\mathrm{d}k_x \right\}. \quad (18)$$

Also, integration over $z$ now extends from $-z_o$ to 0, with $-z_o$ sufficiently far below the prism's base to include the entire overlap zone. Substituting from Eq. (18) into Eq. (10), multiplying through, then time-averaging to eliminate $\pm 2\omega$ terms, we obtain

$$\langle \boldsymbol{F}(\boldsymbol{r},t)\rangle = \tfrac{1}{4}\mathrm{i}\varepsilon_0(\varepsilon-1)\left\{\iint [k_x \mathcal{E}(k_x)\mathcal{E}_x^*(k_x') - k_x' \mathcal{E}_x(k_x)\mathcal{E}^*(k_x')]\{r_p(k_x)\exp[-\mathrm{i}(k_z+k_z')z] + r_p^*(k_x')\exp[\mathrm{i}(k_z+k_z')z]\}\right.$$

$$\times \exp[\mathrm{i}(k_x - k_x')x]\mathrm{d}k_x \mathrm{d}k_x'$$

$$- \iint [k_z \mathcal{E}(k_x)\mathcal{E}_z^*(k_x') + k_z' \mathcal{E}_z(k_x)\mathcal{E}^*(k_x')]\{r_p(k_x)\exp[-\mathrm{i}(k_z+k_z')z] - r_p^*(k_x')\exp[\mathrm{i}(k_z+k_z')z]\}$$

$$\times \exp[\mathrm{i}(k_x - k_x')x]\mathrm{d}k_x \mathrm{d}k_x'$$

$$-\tfrac{1}{4}\mathrm{i}\mu_o\varepsilon_0(\varepsilon-1)\omega \iint [\mathcal{E}(k_x)\times\mathcal{H}^*(k_x') - \mathcal{E}^*(k_x')\times\mathcal{H}(k_x)]\{r_p(k_x)\exp[-\mathrm{i}(k_z+k_z')z] + r_p^*(k_x')\exp[\mathrm{i}(k_z+k_z')z]\}$$

$$\times \exp[\mathrm{i}(k_x - k_x')x]\mathrm{d}k_x \mathrm{d}k_x' \qquad \text{continued …}$$

$$+ \tfrac{1}{4}\mathrm{i}\mu_0\varepsilon_0(\mu-1)\omega \iint [\mathcal{H}(k_x)\times\mathcal{E}^*(k_x') - \mathcal{H}^*(k_x')\times\mathcal{E}(k_x)]\{r_p(k_x)\exp[-\mathrm{i}(k_z+k_z')z] + r_p^*(k_x')\exp[\mathrm{i}(k_z+k_z')z]\}$$
$$\times \exp[\mathrm{i}(k_x-k_x')x]\,\mathrm{d}k_x\,\mathrm{d}k_x'. \tag{19}$$

We now combine similar terms and replace the $\mathcal{H}$-field in terms of the $\mathcal{E}$-field components as given by Eq.(9) to obtain:

$$\langle \boldsymbol{F} \rangle = \tfrac{1}{4}\mathrm{i}\varepsilon_0(\varepsilon-1)\iint \{[k_x\mathcal{E}_x(k_x)\mathcal{E}_x^*(k_x') - k_x'\mathcal{E}_x(k_x)\mathcal{E}_x^*(k_x') - k_z\mathcal{E}_x(k_x)\mathcal{E}_z^*(k_x') - k_z'\mathcal{E}_z(k_x)\mathcal{E}_x^*(k_x')]\hat{x}$$
$$+ [k_x\mathcal{E}_z(k_x)\mathcal{E}_z^*(k_x') - k_x'\mathcal{E}_x(k_x)\mathcal{E}_z^*(k_x') - k_z\mathcal{E}_z(k_x)\mathcal{E}_z^*(k_x') - k_z'\mathcal{E}_z(k_x)\mathcal{E}_z^*(k_x')]\hat{z}\}$$
$$\times r_p(k_x)\exp[-\mathrm{i}(k_z+k_z')z]\exp[\mathrm{i}(k_x-k_x')x]\,\mathrm{d}k_x\,\mathrm{d}k_x'$$
$$+ \tfrac{1}{4}\mathrm{i}\varepsilon_0(\varepsilon-1)\iint\{[k_x\mathcal{E}_x(k_x)\mathcal{E}_x^*(k_x') - k_x'\mathcal{E}_x(k_x)\mathcal{E}_x^*(k_x') + k_z\mathcal{E}_x(k_x)\mathcal{E}_z^*(k_x') + k_z'\mathcal{E}_z(k_x)\mathcal{E}_x^*(k_x')]\hat{x}$$
$$+ [k_x\mathcal{E}_z(k_x)\mathcal{E}_z^*(k_x') - k_x'\mathcal{E}_x(k_x)\mathcal{E}_z^*(k_x') + k_z\mathcal{E}_z(k_x)\mathcal{E}_z^*(k_x') + k_z'\mathcal{E}_z(k_x)\mathcal{E}_z^*(k_x')]\hat{z}\}$$
$$\times r_p^*(k_x')\exp[\mathrm{i}(k_z+k_z')z]\exp[\mathrm{i}(k_x-k_x')x]\,\mathrm{d}k_x\,\mathrm{d}k_x'$$
$$+ \tfrac{1}{4}\mathrm{i}\varepsilon_0[(\mu-\varepsilon)/\mu]\iint[k_x'\mathcal{E}_x(k_x)\mathcal{E}_x^*(k_x') - k_x\mathcal{E}_z(k_x)\mathcal{E}_z^*(k_x') - k_z\mathcal{E}_x(k_x)\mathcal{E}_z^*(k_x') - k_z'\mathcal{E}_z(k_x)\mathcal{E}_x^*(k_x')]\hat{x}$$
$$+ [k_x\mathcal{E}_z(k_x)\mathcal{E}_x^*(k_x') - k_x'\mathcal{E}_x(k_x)\mathcal{E}_z^*(k_x') + k_z\mathcal{E}_x(k_x)\mathcal{E}_z^*(k_x') + k_z'\mathcal{E}_x(k_x)\mathcal{E}_x^*(k_x')]\hat{z}$$
$$\times r_p(k_x)\exp[-\mathrm{i}(k_z+k_z')z]\exp[\mathrm{i}(k_x-k_x')x]\,\mathrm{d}k_x\,\mathrm{d}k_x'$$
$$+ \tfrac{1}{4}\mathrm{i}\varepsilon_0[(\mu-\varepsilon)/\mu]\iint[k_x'\mathcal{E}_x(k_x)\mathcal{E}_z^*(k_x') - k_x\mathcal{E}_z(k_x)\mathcal{E}_x^*(k_x') + k_z\mathcal{E}_x(k_x)\mathcal{E}_z^*(k_x') + k_z'\mathcal{E}_z(k_x)\mathcal{E}_x^*(k_x')]\hat{x}$$
$$+ [k_x\mathcal{E}_z(k_x)\mathcal{E}_x^*(k_x') - k_x'\mathcal{E}_x(k_x)\mathcal{E}_z^*(k_x') - k_z\mathcal{E}_x(k_x)\mathcal{E}_z^*(k_x') - k_z'\mathcal{E}_x(k_x)\mathcal{E}_x^*(k_x')]\hat{z}$$
$$\times r_p^*(k_x')\exp[\mathrm{i}(k_z+k_z')z]\exp[\mathrm{i}(k_x-k_x')x]\,\mathrm{d}k_x\,\mathrm{d}k_x'. \tag{20}$$

Next we use Maxwell's first equation, $k_x\mathcal{E}_x + k_z\mathcal{E}_z = 0$, to further simplify the force density expression, remembering that for the reflected beam the sign of $k_z$ must be reversed.

$$\langle \boldsymbol{F}\rangle = \tfrac{1}{4}\mathrm{i}\varepsilon_0\iint\{\{(\varepsilon-1)(k_x-k_x')\mathcal{E}_x(k_x)\mathcal{E}_x^*(k_x') + \{[(\varepsilon/\mu)-1](k_x-k_x') + [(\varepsilon/\mu)-\varepsilon][(k_z^2/k_x) - (k_z'^2/k_x')]\}\mathcal{E}_z(k_x)\mathcal{E}_z^*(k_x')\}\hat{x}$$
$$- \{[(\varepsilon/\mu)-1](k_z+k_z')\mathcal{E}_x(k_x)\mathcal{E}_x^*(k_x') + \{(\varepsilon-1)(k_z+k_z') - [(\varepsilon/\mu)-\varepsilon][(k_x'k_z/k_x)+(k_xk_z'/k_x')]\}\mathcal{E}_z(k_x)\mathcal{E}_z^*(k_x')\}\hat{z}\}$$
$$\times r_p(k_x)\exp[-\mathrm{i}(k_z+k_z')z]\exp[\mathrm{i}(k_x-k_x')x]\,\mathrm{d}k_x\,\mathrm{d}k_x'$$
$$+ \tfrac{1}{4}\mathrm{i}\varepsilon_0\iint\{\{(\varepsilon-1)(k_x-k_x')\mathcal{E}_x(k_x)\mathcal{E}_x^*(k_x') + \{[(\varepsilon/\mu)-1](k_x-k_x') + [(\varepsilon/\mu)-\varepsilon][(k_z^2/k_x) - (k_z'^2/k_x')]\}\mathcal{E}_z(k_x)\mathcal{E}_z^*(k_x')\}\hat{x}$$
$$+ \{[(\varepsilon/\mu)-1](k_z+k_z')\mathcal{E}_x(k_x)\mathcal{E}_x^*(k_x') + \{(\varepsilon-1)(k_z+k_z') - [(\varepsilon/\mu)-\varepsilon][(k_x'k_z/k_x)+(k_xk_z'/k_x')]\}\mathcal{E}_z(k_x)\mathcal{E}_z^*(k_x')\}\hat{z}\}$$
$$\times r_p^*(k_x')\exp[\mathrm{i}(k_z+k_z')z]\exp[\mathrm{i}(k_x-k_x')x]\,\mathrm{d}k_x\,\mathrm{d}k_x'. \tag{21}$$

To find the total force exerted on the prism of Fig.3 due to the overlap between the incident and reflected beams, we integrate Eq.(21), first from $z=-z_0$ to $0$, then from $x=-\infty$ to $\infty$, to obtain:

$$\int_{-\infty}^{\infty}\int_{-z_0}^{0}\langle \boldsymbol{F}(x,z,t)\rangle\,\mathrm{d}z\,\mathrm{d}x$$
$$= \tfrac{1}{4}\varepsilon_0\iint\{\{(\varepsilon-1)(k_x-k_x')\mathcal{E}_x(k_x)\mathcal{E}_x^*(k_x') + \{[(\varepsilon/\mu)-1](k_x-k_x') + [(\varepsilon/\mu)-\varepsilon][(k_z^2/k_x) - (k_z'^2/k_x')]\}\mathcal{E}_z(k_x)\mathcal{E}_z^*(k_x')\}\hat{x}$$
$$- \{[(\varepsilon/\mu)-1](k_z+k_z')\mathcal{E}_x(k_x)\mathcal{E}_x^*(k_x') + \{(\varepsilon-1)(k_z+k_z') - [(\varepsilon/\mu)-\varepsilon][(k_x'k_z/k_x)+(k_xk_z'/k_x')]\}\mathcal{E}_z(k_x)\mathcal{E}_z^*(k_x')\}\hat{z}\}$$
$$\times r_p(k_x)(k_z+k_z')^{-1}\{\exp[\mathrm{i}(k_z+k_z')z_0] - 1\}\,\delta(k_x-k_x')\,\mathrm{d}k_x\,\mathrm{d}k_x'$$
$$+ \tfrac{1}{4}\varepsilon_0\iint\{\{(\varepsilon-1)(k_x-k_x')\mathcal{E}_x(k_x)\mathcal{E}_x^*(k_x') + \{[(\varepsilon/\mu)-1](k_x-k_x') + [(\varepsilon/\mu)-\varepsilon][(k_z^2/k_x) - (k_z'^2/k_x')]\}\mathcal{E}_z(k_x)\mathcal{E}_z^*(k_x')\}\hat{x}$$

$$+ \{[(\varepsilon/\mu)-1](k_z+k_z')\mathcal{E}_x(k_x)\mathcal{E}_x^*(k_x') + \{(\varepsilon-1)(k_z+k_z')-[(\varepsilon/\mu)-\varepsilon][(k_x'k_z/k_x)+(k_xk_z'/k_x')]\}\mathcal{E}_z(k_x)\mathcal{E}_z^*(k_x')\}\hat{z}\}$$

$$\times r_p^*(k_x')(k_z+k_z')^{-1}\{1-\exp[-\mathrm{i}(k_z+k_z')z_0]\}\delta(k_x-k_x')\mathrm{d}k_x\mathrm{d}k_x'. \tag{22}$$

Upon integration with respect to $k_x'$, the $x$-component of the force vanishes and the $z$-component becomes

$$\int_{-\infty}^{\infty}\int_{-z_0}^{0}<F_z(x,z,t)>\mathrm{d}z\,\mathrm{d}x = \tfrac{1}{4}\varepsilon_0\int\{[(\varepsilon/\mu)-1]|\mathcal{E}_x(k_x)|^2 + [(\varepsilon/\mu)-2\varepsilon+1]|\mathcal{E}_z(k_x)|^2\}$$

$$\times\{r_p(k_x)[1-\exp(2\mathrm{i}k_zz_0)] + r_p^*(k_x)[1-\exp(-2\mathrm{i}k_zz_0)]\}\mathrm{d}k_x. \tag{23}$$

It must be remembered that $\mathcal{E}_z(k_x)$ and $\mathcal{E}_z^*(k_x')$ belong to the incident and reflected beams (not necessarily in that order); therefore, when $k_x$ approaches $k_x'$, these field components acquire equal magnitudes and opposite signs. This explains the sign change of the $\mathcal{E}_z(k_x)\mathcal{E}_z^*(k_x')$ coefficient in going from Eq. (22) to Eq. (23). The exponential factors appearing in Eq. (22) integrate to zero provided that $z_0$ is sufficiently large for these factors to rapidly oscillate within the $k$-space inhabited by the beams. If the bandwidth is sufficiently narrow, $|\mathcal{E}_x|\approx|\mathcal{E}|\cos\theta$, $|\mathcal{E}_z|\approx|\mathcal{E}|\sin\theta$, and the Fresnel reflection coefficient of total internal reflection will be

$$r_p(k_x)\approx r_p(k_{x0}) = (\sqrt{\mu\varepsilon\sin^2\theta-1}+\mathrm{i}\sqrt{\mu/\varepsilon}\cos\theta)/(\sqrt{\mu\varepsilon\sin^2\theta-1}-\mathrm{i}\sqrt{\mu/\varepsilon}\cos\theta). \tag{24}$$

Equation (23) thus reduces to

$$\int_{-\infty}^{\infty}\int_{-z_0}^{0}<F_z(x,z,t)>\mathrm{d}z\,\mathrm{d}x \approx \tfrac{1}{2}\varepsilon_0[(\varepsilon/\mu)-1-2(\varepsilon-1)\sin^2\theta]\,\mathrm{Real}(r_p)\int_{-\infty}^{\infty}|\mathcal{E}(k_x)|^2\mathrm{d}k_x, \tag{25}$$

where

$$\mathrm{Real}(r_p) = \{[(\varepsilon/\mu)+1]-(\varepsilon^2+1)\sin^2\theta\}/\{[(\varepsilon/\mu)-1]-(\varepsilon^2-1)\sin^2\theta\}. \tag{26}$$

Next we compute the force due to the $E$-field discontinuity at the top facet of the prism.

$$E_z(x,z=0^-,t) = \tfrac{1}{2}\left\{\int[1-r_p(k_x)]\mathcal{E}_z(k_x)\exp[\mathrm{i}(k_xx-\omega t)]\mathrm{d}k_x + \int[1-r_p^*(k_x)]\mathcal{E}_z^*(k_x)\exp[-\mathrm{i}(k_xx-\omega t)]\mathrm{d}k_x\right\}, \tag{27a}$$

$$E_z(x,z=0^+,t) = \tfrac{1}{2}\varepsilon\left\{\int[1-r_p(k_x)]\mathcal{E}_z(k_x)\exp[\mathrm{i}(k_xx-\omega t)]\mathrm{d}k_x + \int[1-r_p^*(k_x)]\mathcal{E}_z^*(k_x)\exp[-\mathrm{i}(k_xx-\omega t)]\mathrm{d}k_x\right\}. \tag{27b}$$

We find

$$F_z(x,z=0,t) = \tfrac{1}{2}P_z(x,z=0^-,t)[E_z(x,z=0^+,t)-E_z(x,z=0^-,t)]$$

$$= (1/8)\varepsilon_0(\varepsilon-1)^2\left\{\int[1-r_p(k_x)]\mathcal{E}_z(k_x)\exp[\mathrm{i}(k_xx-\omega t)]\mathrm{d}k_x + \int[1-r_p^*(k_x)]\mathcal{E}_z^*(k_x)\exp[-\mathrm{i}(k_xx-\omega t)]\mathrm{d}k_x\right\}$$

$$\times\left\{\int[1-r_p(k_x)]\mathcal{E}_z(k_x)\exp[\mathrm{i}(k_xx-\omega t)]\mathrm{d}k_x + \int[1-r_p^*(k_x)]\mathcal{E}_z^*(k_x)\exp[-\mathrm{i}(k_xx-\omega t)]\mathrm{d}k_x\right\}. \tag{28}$$

Integrating the above surface force density over $x$ from $-\infty$ to $\infty$ yields

$$\int_{-\infty}^{\infty}<F_z(x,z=0,t)>\mathrm{d}x = \tfrac{1}{4}\varepsilon_0(\varepsilon-1)^2\iint[1-r_p(k_x)][1-r_p^*(k_x')]\mathcal{E}_z(k_x)\mathcal{E}_z^*(k_x')\delta(k_x-k_x')\mathrm{d}k_x\mathrm{d}k_x'$$

$$= \tfrac{1}{4}\varepsilon_0(\varepsilon-1)^2\int_{-\infty}^{\infty}|1-r_p(k_x)|^2|\mathcal{E}_z(k_x)|^2\mathrm{d}k_x. \tag{29}$$

Considering that $|r_p(k_x)|=1$, and that, for a narrowband beam, $|\mathcal{E}_z|\approx|\mathcal{E}|\sin\theta$, the above equation may be written

$$\int_{-\infty}^{\infty}<F_z(x,z=0,t)>\mathrm{d}x \approx \tfrac{1}{2}\varepsilon_0(\varepsilon-1)^2[1-\mathrm{Real}(r_p)]\sin^2\theta\int_{-\infty}^{\infty}|\mathcal{E}(k_x)|^2\mathrm{d}k_x. \tag{30}$$

When Eq. (30) is added to Eq. (25) we find

$$<F_z> = \tfrac{1}{2}\varepsilon_0\{[(\varepsilon/\mu)-1-(\varepsilon^2-1)\sin^2\theta]\mathrm{Real}(r_p)+(\varepsilon-1)^2\sin^2\theta\}\int_{-\infty}^{\infty}|\mathcal{E}(k_x)|^2\mathrm{d}k_x. \tag{31}$$

Substitution from Eq. (26) then yields

$$<F_z> = \tfrac{1}{2}\varepsilon_0[(\varepsilon/\mu)+1-2\varepsilon\sin^2\theta]\int_{-\infty}^{\infty}|\mathcal{E}(k_x)|^2\mathrm{d}k_x. \tag{32}$$

To this force we now add the side-wall forces of the incident and reflected beams given by Eq. (17) and find

$$<F_z> = \tfrac{1}{2}\varepsilon_0[(\varepsilon/\mu)+1]\cos^2\theta \int_{-\infty}^{\infty} |\mathcal{E}(k_x)|^2 dk_x. \tag{33}$$

The above $<F_z>$ must be equal to the net change in the rate of flow of momentum in the free space regions outside the prism plus the force on the anti-reflection (AR) coating layers. The factor $\cos^2\theta$ in Eq. (33) accounts for the direction of the incident beam relative to the $z$-axis, as well as the cross-sectional area of the beam, which is reduced by a factor of $\cos\theta$ relative to the footprint of the beam at the prism's base. As for the coefficient $\tfrac{1}{2}\varepsilon_0[(\varepsilon/\mu)+1]$, recall that, by conservation of energy, the incident $E$-field amplitude in the free space is greater than that inside the prism by a factor of $(\varepsilon/\mu)^{1/4}$. Therefore, the incident momentum flux in the free space is proportional to $\tfrac{1}{2}\varepsilon_0\sqrt{\varepsilon/\mu}$. From Ref. [11], Eq. (13), we know that the force on the AR-coating layer is proportional to $\tfrac{1}{4}\varepsilon_0\sqrt{\varepsilon/\mu}(1-\sqrt{\mu/\varepsilon})(1-\sqrt{\varepsilon/\mu}) = -\tfrac{1}{4}\varepsilon_0(1-\sqrt{\varepsilon/\mu})^2$, the minus sign indicating that the direction of this force is opposed to that of the incident beam. Combining these last two expressions (being careful about the directions of the incident momentum flux and the force acting on the AR-coating) yields a net force of $\tfrac{1}{4}\varepsilon_0[(\varepsilon/\mu)+1]$. The remaining factor of 2 comes from an equal contribution by the reflected beam upon exiting the prism. We have thus shown complete agreement between the force calculated from the generalized Lorentz law as given by Eq. (1a) and the momentum flux of the beams that enter and exit the prism of Fig. 3. A similar calculation based on Eq. (4a) would result in a similar agreement in the end, even though some of the intervening results will be substantially different.

**5. General remarks and conclusions.** The generalized Lorentz law in conjunction with Maxwell's macroscopic equations, Eqs. (2), provide a complete and consistent set of equations that are fully compatible with the laws of conservation of energy and momentum. We have argued that the $E$, $D$, $H$, and $B$ fields should be treated as fundamental, while the polarization and magnetization densities $P$ and $M$ are secondary fields derived from the constitutive relations, Eqs. (3). Two formulations of the generalized Lorentz law of force/torque are admissible, one given by Eqs. (1), the other by Eqs. (4). These formulations are equally acceptable in the sense that they are both consistent with the conservation laws; they predict precisely the same total force/torque on a given body of material, even though the predicted force/torque distributions at surfaces and interfaces and throughout the volumes of material could be drastically different in the two formulations.

Two other postulates that need explicit enunciation in the classical theory of electromagnetism are related to energy and momentum. The first postulate declares that the time rate of change of energy density, $\partial \mathcal{E}/\partial t$, is $E \cdot J_{\text{free}} + E \cdot \partial D/\partial t + H \cdot \partial B/\partial t$. The second postulate establishes the (linear) momentum density of the electromagnetic field as $p_{\text{EM}} = E \times H/c^2$. Once these postulates are accepted, one can easily show that the rate of flow of energy (per unit area per unit time) is the Poynting vector $S = E \times H$, that the linear and angular momenta enter and exit a given medium with the group velocity $V_g$, and that the balance of the incident, reflected and transmitted momenta is experienced by the medium as force and torque exerted by the electromagnetic field. These forces and torques are typically concentrated at the surfaces and interfaces of the material media, at the edges of the beam (i.e., leading and trailing edges as well as the sidewalls), and, more generally, in any region where the intensity gradient is non-vanishing. For instance, in some of the examples discussed in our previous publications [11-20], where the normally-incident beam was fairly uniform, circularly polarized, and had a large diameter (relative to a wavelength), the forces and torques were typically confined to the vicinity of the leading and trailing edges of the beam. Depending on which form of the Lorentz law is used, i.e., Eqs. (1) or Eqs. (4), the forces and torques may appear in one part of the beam or another (e.g., at the leading edge of the beam or at the entrance facet of the medium), the force may be compressive at the beam's side-walls in one formulation and expansive in the other, but in all cases, the total force (and total torque) exerted on the material body will be exactly the same, irrespective of the formulation used.

Going slightly beyond the classical theory, if one assumes that the energy contained in a given volume of space (or material body) is divided into individual packets of $\hbar\omega$, then the electromagnetic momentum corresponding to each such bundle of energy (also known as a photon) will be $\hbar\omega/(n_g c)$, where $n_g$ is the group refractive index of the medium. For circularly-polarized light, the angular momentum of each photon is $\hbar/(n_p n_g)$, where $n_p$ is the phase refractive index. The balance of linear and angular momenta among the incident, reflected, and transmitted beams is always transferred to the medium in the form of force and torque, which are sometimes identified with the "mechanical" momenta of the light beam. When a beam of light enters a material medium from the free space, a fraction of its linear (angular) momentum remains electromagnetic, while the rest is exerted on the medium in the form of mechanical force (torque). When a beam of light emerges from a medium into the free space, its electromagnetic momenta are augmented by additional momenta that are taken away from the medium, resulting in mechanical backlash (i.e., oppositely directed force and torque) on the medium. These results can be proven from first principles (i.e., Maxwell's macroscopic equations in conjunction with the generalized Lorentz law) without resort to any additional hypotheses or approximations.

In addition to the numerous examples discussed in our previous publications [11-20], many cases of practical significance can be analyzed with the analytical methods that have been elaborated in the present paper. The case of a parallel-plate magnetic slab examined in Section 3 not only provides an instance of force computation based on the generalized Lorentz law, it also demonstrates the absence of surface forces when a beam of light enters a transparent, semi-infinite medium at normal incidence. The methods of Section 4, used here to demonstrate conservation of linear momentum in a case involving total internal reflection within a transparent magnetic prism, can be readily extended to compute linear and angular momenta as well as mechanical forces and torques exerted by electromagnetic radiation within dispersive, absorptive, magnetic dielectrics.

**Acknowledgements.** This work has been supported by the Air Force Office of Scientific Research (AFOSR) under contract number FA 9550−04−1−0213. The author is grateful to Ewan M. Wright for many helpful discussions.